\documentclass[%
preprint,
 amsmath,amssymb,
 aip,
 apl,
 numerical,
 floats,
]{revtex4-1}
\usepackage{graphicx,color}% Include figure files
\usepackage{dcolumn}% Align table columns on decimal point
\usepackage{bm}% bold math
\usepackage{epsf}
\usepackage{subfig}
\usepackage{ulem}

\begin{document}
%\preprint{}

\newcommand{\av}[1]{{\color{red}$\clubsuit$#1}}
\newcommand{\ml}[1]{{\color{green}$\spadesuit$#1}}

\title{\textit{Ab initio} calculation of Effective Work Functions for a TiN/HfO$_2$/SiO$_2$/Si transistor stack}

\author{Pierre-Yves Prodhomme}
\affiliation{CEA-LETI, MINATEC Campus, 17 rue des Martyrs, 38054 Grenoble, France}
\affiliation{FOTON-INSA Laboratory, UMR 6082 au CNRS, INSA de Rennes, 20 Avenue des Buttes de Coësmes,
CS 14315, 35043 Rennes Cedex, France}
\author{Fabien Fontaine-Vive}
\affiliation{CEA-LETI, MINATEC Campus, 17 rue des Martyrs, 38054 Grenoble, France}
\author{Abram Van Der Geest}
\affiliation{CEA-LETI, MINATEC Campus, 17 rue des Martyrs, 38054 Grenoble, France}
\author{Philippe Blaise}
\affiliation{CEA-LETI, MINATEC Campus, 17 rue des Martyrs, 38054 Grenoble, France}
\author{Jacky Even}
\affiliation{FOTON-INSA Laboratory, UMR 6082 au CNRS, INSA de Rennes, 20 Avenue des Buttes de Coësmes,
CS 14315, 35043 Rennes Cedex, France}

\begin{abstract}
\textit{Ab initio} techniques are used to calculate the effective work function ($\mathrm{W_{eff}}$) of a TiN/HfO$_2$/SiO$_2$/Si stack representing a metal-oxide-semiconductor (MOS) transistor gate taking into account first order many body effects. The required band offsets were calculated at each interface varying its composition. Finally the transitivity of LDA calculated bulk band lineups were used and completed by MBPT bulk corrections for the terminating materials (Si and TiN) of the MOS stack. With these corrections the ab initio calculations predict a $\mathrm{W_{eff}}$ of a TiN metal gate on HfO$_2$ to be close to 5.0 eV.

\end{abstract}

%\pacs{30}% PACS, the Physics and Astronomy
                             % Classification Scheme.
%\keywords{Suggested keywords}%Use showkeys class option if keyword
                              %display desired
\maketitle

\pagebreak

%\subsection{Aims}
Reducing the scale of the Metal-Oxide-Semiconductor (MOS) transistor has led the semiconductor industry to major changes in the MOS stack.  First, the SiO$_2$ oxide has been replaced by a HfO$_2$ layer on top of a SiO$_2$ thin film, while TiN is a leading candidate to replace poly-silicon in the gate for the next generation of CMOS transistor (22 nm and sub-22 nm). However, the control of the threshold voltage remains one of the major issues.  The effective work function ($\mathrm{W_{eff}}$) of a metal in a MOS structure is one of the key properties governing this threshold voltage. The aim of this paper is to construct a milestone for the comparison between an ideal system and available experimental data.\

During electrical capacitance-voltage (C-V) measurements $\mathrm{W_{eff}}$ is usually estimated through:
\begin{equation}\label{eq1}
\mathrm{W_{eff}}=\phi_{\mathrm{ms}}+\chi(Si)+\zeta(Si).
\end{equation}
Here $\chi(Si)$ is the electronic affinity of silicon,  $\zeta(Si)$ is the difference between the conduction band and the Fermi energy of doped silicon, and $\phi_{ms}$ is the difference between the metal Fermi energy and the semiconductor Fermi energy.  $\phi_{\mathrm{ms}}$ is extracted from the flat-band voltage through modeling of the %\sout{default}
defect's charge distribution along the oxide\cite{Atsushi_2006}.

The valence band offset (VBO) between the metal and the semiconductor (VBO$_{\mathrm{ms}}$) is related to $\phi_{\mathrm{ms}}$ through $\mathrm{VBO_{ms}} +E_g(Si)=\phi_{\mathrm{ms}}+\zeta(Si)$, where E$_g$(Si)(1.1 eV) is the electronic gap of silicon. Combining this relation with equation \ref{eq1} and considering the well-accepted value of $\chi(Si)$ (4.1 eV), the
\textit{ab initio} $\mathrm{W_{eff}}$ of a given metal can be evaluated through the equation
\begin{equation}
\mathrm{W_{eff}^{ab\ initio}}=\mathrm{VBO_{ms}^{ab\ initio}}+\chi(Si)+E_g(Si).
\end{equation}
Then one realizes that the \textit{ab initio} quantity of primary importance is $\mathrm{VBO_{ms}}$.\

Evaluating the $\mathrm{VBO_{ms}}$ from first principles calculations is a challenge. On the one hand the recent introduction of hafnium oxide requires the accurate treatment of the complex chemistry of the interfaces. On the other hand, in order to reach a sufficient precision on electronic energies ($\pm$ $0.1$ eV), one needs to compute the eigenvalues in Many Body Perturbation Theory (MBPT).  However, this is out of reach for the treatment of a complete MOS stack within the GW approximation (GWA, where G stands for the Green's function and W for the screened potential). Fortunately, the interface's potential drop can be well estimated within the Density Functional Theory (DFT) \cite{Shaltaf_PRL_2008}, and coupled with the GW eigenvalues for the valence band states calculated only for each bulk material. These eigenvalues are calculated in reference to the electrostatic potential average\cite{Junquera_2007} of each bulk.   This reference potential is then aligned with the potential drop at the interface producing the band offset\cite{Shaltaf_PRL_2008}. This technique is generally applicable for a sufficient material thickness (1 to 2 nm thickness for each slab) such that each material approaches a bulk state away from the interface.

In this manner, the first order implementation of GWA, called G$_0$W$_0$, has been applied to correct the band alignments along III-V heterojunctions.  These corrections slightly improved  the DFT results in the local density approximation (LDA), which were already in pretty good agreement with experiments\cite{Zhang_90,VdW&M}.  Recently, this \textit{ab initio} method has been applied to assess the band alignment at a junction between an oxide (SiO$_2$, ZrO$_2$ or HfO$_2$) and a semiconductor (Si) resulting in a close agreement with experiment unlike the LDA results \cite{Shaltaf_PRL_2008,Gruning_PRB_2010}. In this paper we use the same approach for each interface of a complete TiN / HfO$_2$ / SiO$_2$ / Si stack.

The potential drops at each of the interfaces have been computed within DFT/LDA using the atomic orbital SIESTA software\cite{Siesta_02} whereas the GWA computations have been carried out with the plane-waves ABINIT software\cite{Abinit_2005}. To be consistent, identical Troullier-Martins pseudopotentials \cite{Troullier1991} for each atomic species have been used for Abinit and Siesta and were generated using the Fhi98pp and Atom programs \cite{CompPhysCom_Fuchs_1999,Siesta_02} in the local density approximation (LDA)\cite{Ceperley-Alder}. Siesta calculations have been performed using a polarized double zeta, DZP, basis with an energy shift of 50 meV and a Meshcutoff of at least 100 Hartrees.
A simulated anneal has also been carried out within Siesta as an heuristic method to relax the forces for each stack: from 1000K to 100K in 1 picosecond, followed by a conjugated gradient method to reduce the forces to less than $0.07$ eV/\AA{}.

All GWA calculations were performed at the G$_0$W$_0$ level.  The quasiparticles energies were carefully converged with respect to the energy cut-offs, the number of k-points in the Brillouin zone, and the number of bands used to compute the dielectric function and the self-energy. Thus a numerical accuracy of $\pm0.1$ eV on the quasiparticles energies was reached. The Plasmon-Pole model proposed by Godby and Needs\cite{Godby-Needs_89} was used to describe the dynamic dependence of the ''screening'' function.

%\subsection{First part : Si/SiO$_2$ stack}

Two types of Si/SiO$_2$ interfaces have each been simulated, one oxygen poor and one oxygen rich. These supercells contained a 22.2 \AA{} long slab of $\beta$-cristobalite SiO$_2$ and a 24.3 \AA{} long slab of diamond Si, both oriented with the (001) plane at the interface. The latter slab was compressed by 3\% in the [100] and [010] directions. The simulated LDA value of the $\mathrm{VBO_1^{LDA}}$ is $+3.0$ eV ($+2.8$ eV) for the oxygen rich (poor) interface using the sign convention where VBO$>$0 when the Valence Band Maximum (VBM) of the material on the left is above the VBM of the material on the right. The G$_0$W$_0$ corrections to the VBMs computed for each material are $\delta\epsilon_v^{GW-LDA}(Si)=-0.6$ eV and $\delta\epsilon_v^{GW-LDA}(SiO_2)=-1.9$ eV, which lead to a $\mathrm{VBO_1^{GW}}$ of $+4.2\pm0.1 eV$. These results are in good agreement with the results published by Shaltaf \textit{et al.} \cite{Shaltaf_PRL_2008} ($4.1$ eV calculated within GWA starting from GGA electronic structures), and a little lower than experimental data ranging from $4.3$ eV to $4.5$ eV estimated by X-ray photoemission spectroscopy \cite{sayan_2002,alay_1997}.

\begin{figure}[hbtp]
\centering
{\includegraphics[width=100mm]{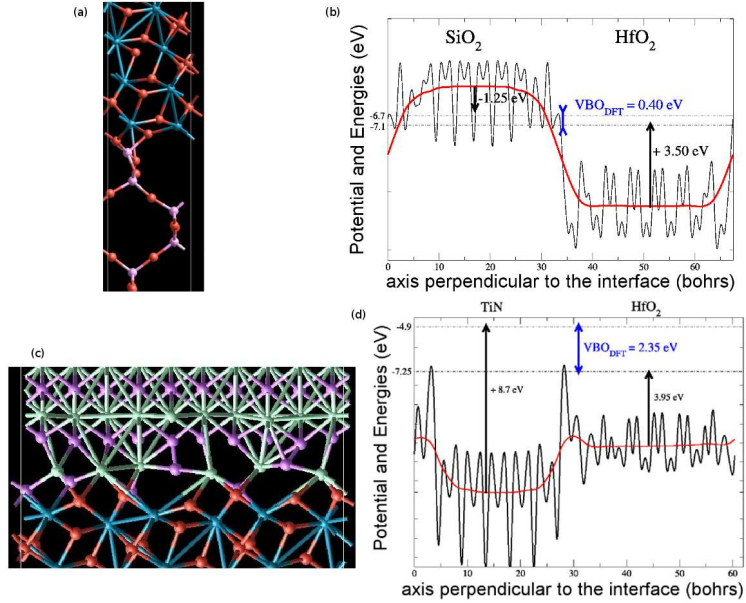}}
\caption{\label{fig:figure}O rich SiO$_2$/HfO$_2$ interface (a) and nitridized TiN/HfO$_2$ interface (c): N in purple, Ti in turquoise, Hf in blue, Si in pink and O in red. Potential projected along the HfO$_2$/SiO$_2$(b) and the TiN/HfO$_2$(d) superlattices (black line) and smoothed by a double convolution (red line). The black arrows indicate the shifts of the eigenenergies with respect to the averaged potential in the bulk. The blue arrow indicates the $\mathrm{VBO^{LDA}}$.}
\end{figure}

%\subsection{Second part : SiO$_2$/HfO$_2$ stack}
The SiO$_2$/HfO$_2$ interface is the second interface of the complete stack (figure \ref{fig:figure}) considered here. The HfO$_2$ slab was simulated in its monoclinic structure with P2$_1$/\textit{c} symmetry, which is its stablest phase at ambient pressure and temperature \cite{Leger_1993,Leger_1999} with the (001) plane at the interface. The SiO$_2$ slab was again $\beta$-cristobalite and oriented with the (001) plane at the interface.  The HfO$_2$ was stretched by 3.7\% in the [100] direction and by 2.1\% in the [010] direction to match the $5.23\times5.23$ \AA{}$^2$ square interface of the SiO$_2$, which agrees with the other simulations in literature of SiO$_2$ in the stack \cite{Shaltaf_PRL_2008, Sharia_PRB_2007}. The supercell contained seven HfO$_2$ layers and ten SiO$_2$ tetrahedron.
At the interface after relaxation, 3 oxygens were each three times coordinated with both Si and Hf.  The potential along the stack shown in figure \ref{fig:figure} exhibits a large potential drop of 5.2 eV, which is expected since the electric permittivity and the average valence electronic density are much higher in hafnia than in silica.
At the LDA level the $\mathrm{VBO_2^{LDA}}$ is $+0.4$ eV, which is qualitatively wrong with respect to XPS measurements \cite{sayan_2002}, i.e. the VBM of SiO$_2$ is measured below the VBM of HfO$_2$. Some DFT calculations of the HfO$_2$/SiO$_2$ interface at the LDA level have also been performed by Sharia \textit{et al.} showing that the valence band offset is highly dependent on the interface stoichiometry \cite{Sharia_PRB_2007}.
Using our G$_0$W$_0$ correction for SiO$_2$ ($-1.9$ eV) and for HfO$_2$ ($-0.5$ eV close to the value of Gr\"uning \textit{et al.}\cite{Gruning_PRB_2010} of $-0.4$ eV) the LDA band offset is reversed producing a $\mathrm{VBO_2^{GW}}$ of $-1.0$ eV. This is qualitatively and quantitatively in good agreement with XPS measurements \cite{sayan_2002} of $1.05\pm0.1$ eV.
%\subsection{Third part : TiN/HfO$_2$ stack}

The TiN/HfO$_2$ interface is the final interface of the stack.  Constructing this interface requires a specific procedure because of the large lattice mismatch. First,  a slab of 6 layers of monoclinic HfO$_2$ surrounded by vacuum oriented in the direction perpendicular to the (001) plane containing three unit cells in the [100] direction and one unit cell in the [010] direction was generated.  During molecular dynamic simulations several TiN layers were successively added to the surface.   We found that the positions of the interfacial atoms from TiN were influenced by the underlying HfO$_2$ slab arranging themselves in almost the same structure as the oxide (at $\pm$ $0.3$\AA{} of the atom sites).  Thanks to this observation when we constructed our TiN/HfO$_2$ stack model the Ti atoms at the interface were placed in the sites of the hafnium and the nitrogens in the oxygen sites.  The stack was finalized by adding a face-centered-cubic TiN slab oriented with the (111) surface plane on top of the interfacial TiN layer, which minimizes the mismatch between the slabs. Indeed, it has been pointed out experimentally that the (111) growth direction lowers the film strain \cite{oh_93, je_97}.
In our case the TiN is $1.8$\% stretched out in the [$1\bar10$] direction and $0.5$\% compressed in the [$11\bar2$] direction, and contained 6 TiN layers expanded by 1\% in the [111] direction in order to keep the average electronic density of the TiN bulk.  Two types of interfaces were computed: one ``nitridized'' with oxygen sites of HfO$_2$ substituted by N at the interface (figure \ref{fig:figure}), the second ``oxidized'', with six O replacing the six N at the interface.
The $\mathrm{VBO^{LDA}_3}$ obtained for these two interfaces are $-2.4$ eV (figure \ref{fig:figure}) and $-2.5$ eV respectively. Previous DFT calculations have been done on TiN/HfO$_2$ stacks but cannot be compared with ours since they were computed in the generalized gradient approximation\cite{Fonseca_PRB_2006}.
The quasiparticle correction to the LDA Fermi energy of TiN was calculated to be $\delta\epsilon_v^{GW-LDA}(TiN)=+0.4$ eV. %\av{do we want to put the HfO2/TiN "VBO"$_{GW}$ in as well?  It feels odd to not include this here. }

\begin{table*}
\centering
\caption{Valence band (VBO) and conduction band (CBO) offsets with the resulting effective work function. The sign of the band offset is positive when the VBM of the material on the left is above the VBM of the material on the right.}
\begin{tabular}{cr@{.}lr@{.}lr@{.}lr@{.}lr@{.}lr@{.}lr@{.}l}
\hline
\hline
\multicolumn{1}{c}{\rule[0mm]{0mm}{0mm}interfaces} & \multicolumn{4}{c}{Si/SiO$_2$} & \multicolumn{2}{c}{SiO$_2$/HfO$_2$} & \multicolumn{4}{c}{HfO$_2$/TiN} & \multicolumn{2}{c}{Si/TiN}  & \multicolumn{2}{c}{W$_{\mathrm{eff}}$(TiN) }\\
\multicolumn{1}{c}{\rule[0mm]{0mm}{0mm}} & \multicolumn{2}{c}{O rich} & \multicolumn{2}{c}{O poor} & \multicolumn{2}{c}{} & \multicolumn{2}{c}{nitridized} & \multicolumn{2}{c}{oxidized} & \multicolumn{2}{c}{} & \multicolumn{2}{c}{}\\
\hline
VBO(LDA)    & +3&0 & +2&8 & +0&4 & -2&4 & -2&5 & \multicolumn{2}{c}{[+0.7, +1.0]} & \multicolumn{2}{c}{[5.9, 6.2]}\\
VBO(G$_0$W$_0$) & +4&3 & +4&1 & -1&0 & -3&3 & -3&4 & \multicolumn{2}{c}{[-0.3,  0.0]} & \multicolumn{2}{c}{[4.9, 5.2]}\\
CBO(G$_0$W$_0$) & -3&2 & -3&4 & +1&6 & +2&7 & +2&6 & \multicolumn{2}{c}{[+0.8, +1.1]} & \multicolumn{2}{c}{}\\
\hline
\hline
\end{tabular}
\label{tabrecap}
\end{table*}
\normalsize

%\subsection{Conclusion}
The valence band alignments along the whole stack at the LDA and G$_0$W$_0$ levels where,
\begin{equation}
\mathrm{VBO_{ms}}=\mathrm{VBO_1}+\mathrm{VBO_2}+\mathrm{VBO_3},
\end{equation}
are depicted in table \ref{tabrecap}. Also included in table \ref{tabrecap} are the conduction band alignments calculated by using our G$_0$W$_0$ band gaps for Si (1.1 eV), SiO$_2$ (8.6 eV), and HfO$_2$ (6.0 eV).  The $\mathrm{VBO_{ms}^{LDA}}$ ranges between $-1.0$ eV and $-0.7$ eV depending on the chemistry of the interfaces which would correspond to a work function of between $6.2$ eV and $5.9$ eV. This is in qualitative disagreement with the experimental $\mathrm{W_{eff}}$ which is usually equal to or below 5 eV.  It is worthy of note that in order to calculate the VBO between the two materials on the extremities of the stack one only needs to compute the GW energies for  the top and the bottom slabs because for the intermediate slabs the corrections vanish yielding,
\begin{equation}
\mathrm{VBO_{ms}^{GW}}=\mathrm{VBO_{ms}^{LDA}}+\delta\epsilon_v^{GW-LDA}(Si)+\delta\epsilon_v^{GW-LDA}(TiN).
\end{equation}
$\mathrm{VBO_{ms}^{GW}}$ ranges from $0.0$ eV to $0.3$ eV depending on the chemistry of the interfaces, and one finally obtains an \textit{ab initio} $\mathrm{W_{eff}}$ evaluated in the MBPT framework ranging from $5.2$ eV to $4.9$ eV  (i.e. P+ in jargon of the semiconductors industry). The $\mathrm{W_{eff}}$  for (111) oriented TiN is in agreement with those measured during cold growth process which mostly does not modify the TiN stoichiometry \cite{Atsushi_2006,wu:113510}. We also point out that the GW corrections applied with a coherent methodology are mandatory in order to obtain qualitatively correct results for the evaluations of VBOs and $\mathrm{W_{eff}}$ in a MOS stack.

We think that our results can be used to adjust the parameters of the model used to simulate the whole MOS device. Secondly, by comparing our 'ideal' model system (no defects and a perfect stoichiometry) with the dispersion of the experimental values (from $4.2$ to $4.9$ eV) assessed by C-V measurements \cite{Charbonnier_TED_2010,Atsushi_2006} we think that our values can be used as a reference to study the impact of the fine chemistry in such a gate stack. Finally, in this paper we show that the $\mathrm{W_{eff}}$ is an accessible and reliable \textit{ab initio} quantity and that this type of calculation is promising to help the understanding of the $\mathrm{W_{eff}}$ for nanodevices in microelectronics.

%\bibliography{biblio_v6}

\begin{thebibliography}{22}
\expandafter\ifx\csname natexlab\endcsname\relax\def\natexlab#1{#1}\fi
\expandafter\ifx\csname bibnamefont\endcsname\relax
  \def\bibnamefont#1{#1}\fi
\expandafter\ifx\csname bibfnamefont\endcsname\relax
  \def\bibfnamefont#1{#1}\fi
\expandafter\ifx\csname citenamefont\endcsname\relax
  \def\citenamefont#1{#1}\fi
\expandafter\ifx\csname url\endcsname\relax
  \def\url#1{\texttt{#1}}\fi
\expandafter\ifx\csname urlprefix\endcsname\relax\def\urlprefix{URL }\fi
\providecommand{\bibinfo}[2]{#2}
\providecommand{\eprint}[2][]{\url{#2}}

\bibitem[{\citenamefont{Kuriyama et~al.}(Sept. 2006)\citenamefont{Kuriyama,
  Faynot, Brevard, Tozzo, Clerc, Deleonibus, Mitard, Vidal, Cristoloveanu, and
  Iwai}}]{Atsushi_2006}
\bibinfo{author}{\bibfnamefont{A.}~\bibnamefont{Kuriyama}},
  \bibinfo{author}{\bibfnamefont{O.}~\bibnamefont{Faynot}},
  \bibinfo{author}{\bibfnamefont{L.}~\bibnamefont{Brevard}},
  \bibinfo{author}{\bibfnamefont{A.}~\bibnamefont{Tozzo}},
  \bibinfo{author}{\bibfnamefont{L.}~\bibnamefont{Clerc}},
  \bibinfo{author}{\bibfnamefont{S.}~\bibnamefont{Deleonibus}},
  \bibinfo{author}{\bibfnamefont{J.}~\bibnamefont{Mitard}},
  \bibinfo{author}{\bibfnamefont{V.}~\bibnamefont{Vidal}},
  \bibinfo{author}{\bibfnamefont{S.}~\bibnamefont{Cristoloveanu}},
  \bibnamefont{and} \bibinfo{author}{\bibfnamefont{H.}~\bibnamefont{Iwai}},
  \bibinfo{journal}{Solid-State Device Research Conference, 2006. ESSDERC 2006.
  Proceeding of the 36th European} pp. \bibinfo{pages}{109--112}
  (\bibinfo{year}{Sept. 2006}), ISSN \bibinfo{issn}{1930-8876}.

\bibitem[{\citenamefont{Shaltaf et~al.}(2008)\citenamefont{Shaltaf, Rignanese,
  Gonze, Giustino, and Pasquarello}}]{Shaltaf_PRL_2008}
\bibinfo{author}{\bibfnamefont{R.}~\bibnamefont{Shaltaf}},
  \bibinfo{author}{\bibfnamefont{G.-M.} \bibnamefont{Rignanese}},
  \bibinfo{author}{\bibfnamefont{X.}~\bibnamefont{Gonze}},
  \bibinfo{author}{\bibfnamefont{F.}~\bibnamefont{Giustino}}, \bibnamefont{and}
  \bibinfo{author}{\bibfnamefont{A.}~\bibnamefont{Pasquarello}},
  \bibinfo{journal}{Phys. Rev. Lett.} \textbf{\bibinfo{volume}{100}},
  \bibinfo{pages}{186401} (\bibinfo{year}{2008}).

\bibitem[{\citenamefont{Junquera et~al.}(2007)\citenamefont{Junquera, Cohen,
  and Rabe}}]{Junquera_2007}
\bibinfo{author}{\bibfnamefont{J.}~\bibnamefont{Junquera}},
  \bibinfo{author}{\bibfnamefont{M.~H.} \bibnamefont{Cohen}}, \bibnamefont{and}
  \bibinfo{author}{\bibfnamefont{K.~M.} \bibnamefont{Rabe}},
  \bibinfo{journal}{Journal of Physics: Condensed Matter}
  \textbf{\bibinfo{volume}{19}}, \bibinfo{pages}{213203 (34pp)}
  (\bibinfo{year}{2007}).

\bibitem[{\citenamefont{Zhang et~al.}(1990)\citenamefont{Zhang, Cohen, Louie,
  Tomanek, and Hybertsen}}]{Zhang_90}
\bibinfo{author}{\bibfnamefont{S.}~\bibnamefont{Zhang}},
  \bibinfo{author}{\bibfnamefont{M.~L.} \bibnamefont{Cohen}},
  \bibinfo{author}{\bibfnamefont{S.~G.} \bibnamefont{Louie}},
  \bibinfo{author}{\bibfnamefont{D.}~\bibnamefont{Tomanek}}, \bibnamefont{and}
  \bibinfo{author}{\bibfnamefont{M.~S.} \bibnamefont{Hybertsen}},
  \bibinfo{journal}{Physical Review B} \textbf{\bibinfo{volume}{41}},
  \bibinfo{pages}{10058} (\bibinfo{year}{1990}).

\bibitem[{\citenamefont{Van~de Walle and Martin}(1986)}]{VdW&M}
\bibinfo{author}{\bibfnamefont{C.~G.} \bibnamefont{Van~de Walle}}
  \bibnamefont{and} \bibinfo{author}{\bibfnamefont{R.~M.}
  \bibnamefont{Martin}}, \bibinfo{journal}{Phys. Rev. B}
  \textbf{\bibinfo{volume}{34}}, \bibinfo{pages}{5621} (\bibinfo{year}{1986}).

\bibitem[{\citenamefont{Gr\"uning et~al.}(2010)\citenamefont{Gr\"uning,
  Shaltaf, and Rignanese}}]{Gruning_PRB_2010}
\bibinfo{author}{\bibfnamefont{M.}~\bibnamefont{Gr\"uning}},
  \bibinfo{author}{\bibfnamefont{R.}~\bibnamefont{Shaltaf}}, \bibnamefont{and}
  \bibinfo{author}{\bibfnamefont{G.-M.} \bibnamefont{Rignanese}},
  \bibinfo{journal}{Phys. Rev. B} \textbf{\bibinfo{volume}{81}},
  \bibinfo{pages}{035330} (\bibinfo{year}{2010}).

\bibitem[{\citenamefont{Soler et~al.}(2002)\citenamefont{Soler, Artacho, Gale,
  Garcia, Junquera, Ordejon, and Sanchez-Portal}}]{Siesta_02}
\bibinfo{author}{\bibfnamefont{J.}~\bibnamefont{Soler}},
  \bibinfo{author}{\bibfnamefont{O.}~\bibnamefont{Artacho}},
  \bibinfo{author}{\bibfnamefont{J.}~\bibnamefont{Gale}},
  \bibinfo{author}{\bibfnamefont{A.}~\bibnamefont{Garcia}},
  \bibinfo{author}{\bibfnamefont{J.}~\bibnamefont{Junquera}},
  \bibinfo{author}{\bibfnamefont{P.}~\bibnamefont{Ordejon}}, \bibnamefont{and}
  \bibinfo{author}{\bibfnamefont{D.}~\bibnamefont{Sanchez-Portal}},
  \bibinfo{journal}{Journal of Physics: Condensed Matter}
  \textbf{\bibinfo{volume}{14}}, \bibinfo{pages}{2745} (\bibinfo{year}{2002}).

\bibitem[{\citenamefont{Gonze et~al.}(2005)\citenamefont{Gonze, Rignanese,
  Verstraete, Beuken, Pouillon, Caracas, Jollet, Detraux, Fuchs, Sindic
  et~al.}}]{Abinit_2005}
\bibinfo{author}{\bibfnamefont{X.}~\bibnamefont{Gonze}},
  \bibinfo{author}{\bibfnamefont{G.}~\bibnamefont{Rignanese}},
  \bibinfo{author}{\bibfnamefont{M.}~\bibnamefont{Verstraete}},
  \bibinfo{author}{\bibfnamefont{J.}~\bibnamefont{Beuken}},
  \bibinfo{author}{\bibfnamefont{Y.}~\bibnamefont{Pouillon}},
  \bibinfo{author}{\bibfnamefont{R.}~\bibnamefont{Caracas}},
  \bibinfo{author}{\bibfnamefont{F.}~\bibnamefont{Jollet}},
  \bibinfo{author}{\bibfnamefont{F.}~\bibnamefont{Detraux}},
  \bibinfo{author}{\bibfnamefont{M.}~\bibnamefont{Fuchs}},
  \bibinfo{author}{\bibfnamefont{L.}~\bibnamefont{Sindic}},
  \bibnamefont{et~al.}, \bibinfo{journal}{Zeit. Kristallogr.}
  \textbf{\bibinfo{volume}{220}}, \bibinfo{pages}{558} (\bibinfo{year}{2005}).

\bibitem[{\citenamefont{Troullier and Martins}(1991)}]{Troullier1991}
\bibinfo{author}{\bibfnamefont{N.}~\bibnamefont{Troullier}} \bibnamefont{and}
  \bibinfo{author}{\bibfnamefont{J.~L.} \bibnamefont{Martins}},
  \bibinfo{journal}{Phys. Rev. B} \textbf{\bibinfo{volume}{43}},
  \bibinfo{pages}{1993} (\bibinfo{year}{1991}).

\bibitem[{\citenamefont{Fuchs and Scheffler}(1999)}]{CompPhysCom_Fuchs_1999}
\bibinfo{author}{\bibfnamefont{M.}~\bibnamefont{Fuchs}} \bibnamefont{and}
  \bibinfo{author}{\bibfnamefont{M.}~\bibnamefont{Scheffler}},
  \bibinfo{journal}{Computer Physics Communications}
  \textbf{\bibinfo{volume}{119}}, \bibinfo{pages}{67 } (\bibinfo{year}{1999}).

\bibitem[{\citenamefont{Ceperley and Alder}(1980)}]{Ceperley-Alder}
\bibinfo{author}{\bibfnamefont{D.~M.} \bibnamefont{Ceperley}} \bibnamefont{and}
  \bibinfo{author}{\bibfnamefont{B.~J.} \bibnamefont{Alder}},
  \bibinfo{journal}{Phys. Rev. Lett.} \textbf{\bibinfo{volume}{45}},
  \bibinfo{pages}{566} (\bibinfo{year}{1980}).

\bibitem[{\citenamefont{Godby and Needs}(1989)}]{Godby-Needs_89}
\bibinfo{author}{\bibfnamefont{R.~W.} \bibnamefont{Godby}} \bibnamefont{and}
  \bibinfo{author}{\bibfnamefont{R.~J.} \bibnamefont{Needs}},
  \bibinfo{journal}{Phys. Rev. Lett.} \textbf{\bibinfo{volume}{62}},
  \bibinfo{pages}{1169} (\bibinfo{year}{1989}).

\bibitem[{\citenamefont{Sayan et~al.}(2002)\citenamefont{Sayan, Garfunkel, and
  Suzer}}]{sayan_2002}
\bibinfo{author}{\bibfnamefont{S.}~\bibnamefont{Sayan}},
  \bibinfo{author}{\bibfnamefont{E.}~\bibnamefont{Garfunkel}},
  \bibnamefont{and} \bibinfo{author}{\bibfnamefont{S.}~\bibnamefont{Suzer}},
  \bibinfo{journal}{Applied Physics Letters} \textbf{\bibinfo{volume}{80}},
  \bibinfo{pages}{2135} (\bibinfo{year}{2002}).

\bibitem[{\citenamefont{Alay and Hirose}(1997)}]{alay_1997}
\bibinfo{author}{\bibfnamefont{J.~L.} \bibnamefont{Alay}} \bibnamefont{and}
  \bibinfo{author}{\bibfnamefont{M.}~\bibnamefont{Hirose}},
  \bibinfo{journal}{Journal of Applied Physics} \textbf{\bibinfo{volume}{81}},
  \bibinfo{pages}{1606} (\bibinfo{year}{1997}).

\bibitem[{\citenamefont{Leger et~al.}(1993)\citenamefont{Leger, Atouf,
  Tomaszewski, and Pereira}}]{Leger_1993}
\bibinfo{author}{\bibfnamefont{J.}~\bibnamefont{Leger}},
  \bibinfo{author}{\bibfnamefont{A.}~\bibnamefont{Atouf}},
  \bibinfo{author}{\bibfnamefont{P.}~\bibnamefont{Tomaszewski}},
  \bibnamefont{and} \bibinfo{author}{\bibfnamefont{A.}~\bibnamefont{Pereira}},
  \bibinfo{journal}{Phys. Rev. B} \textbf{\bibinfo{volume}{48}},
  \bibinfo{pages}{93} (\bibinfo{year}{1993}).

\bibitem[{\citenamefont{Lowther et~al.}(1999)\citenamefont{Lowther, dewhurst,
  Leger, and Haines}}]{Leger_1999}
\bibinfo{author}{\bibfnamefont{J.}~\bibnamefont{Lowther}},
  \bibinfo{author}{\bibfnamefont{J.}~\bibnamefont{dewhurst}},
  \bibinfo{author}{\bibfnamefont{J.}~\bibnamefont{Leger}}, \bibnamefont{and}
  \bibinfo{author}{\bibfnamefont{J.}~\bibnamefont{Haines}},
  \bibinfo{journal}{Phys. Rev. B} \textbf{\bibinfo{volume}{60}},
  \bibinfo{pages}{14485} (\bibinfo{year}{1999}).

\bibitem[{\citenamefont{Sharia et~al.}(2007)\citenamefont{Sharia, Demkov,
  Bersuker, and Lee}}]{Sharia_PRB_2007}
\bibinfo{author}{\bibfnamefont{O.}~\bibnamefont{Sharia}},
  \bibinfo{author}{\bibfnamefont{A.~A.} \bibnamefont{Demkov}},
  \bibinfo{author}{\bibfnamefont{G.}~\bibnamefont{Bersuker}}, \bibnamefont{and}
  \bibinfo{author}{\bibfnamefont{B.~H.} \bibnamefont{Lee}},
  \bibinfo{journal}{Phys. Rev. B} \textbf{\bibinfo{volume}{75}},
  \bibinfo{pages}{035306} (\bibinfo{year}{2007}).

\bibitem[{\citenamefont{Oh and Je}(1993)}]{oh_93}
\bibinfo{author}{\bibfnamefont{U.~C.} \bibnamefont{Oh}} \bibnamefont{and}
  \bibinfo{author}{\bibfnamefont{J.~H.} \bibnamefont{Je}},
  \bibinfo{journal}{Journal of Applied Physics} \textbf{\bibinfo{volume}{74}},
  \bibinfo{pages}{1692} (\bibinfo{year}{1993}).

\bibitem[{\citenamefont{Je et~al.}(1997)\citenamefont{Je, Noh, Kim, and
  Liang}}]{je_97}
\bibinfo{author}{\bibfnamefont{J.~H.} \bibnamefont{Je}},
  \bibinfo{author}{\bibfnamefont{D.~Y.} \bibnamefont{Noh}},
  \bibinfo{author}{\bibfnamefont{H.~K.} \bibnamefont{Kim}}, \bibnamefont{and}
  \bibinfo{author}{\bibfnamefont{K.~S.} \bibnamefont{Liang}},
  \bibinfo{journal}{Journal of Applied Physics} \textbf{\bibinfo{volume}{81}},
  \bibinfo{pages}{6126} (\bibinfo{year}{1997}).

\bibitem[{\citenamefont{Fonseca and Knizhnik}(2006)}]{Fonseca_PRB_2006}
\bibinfo{author}{\bibfnamefont{L.~R.~C.} \bibnamefont{Fonseca}}
  \bibnamefont{and} \bibinfo{author}{\bibfnamefont{A.~A.}
  \bibnamefont{Knizhnik}}, \bibinfo{journal}{Phys. Rev. B}
  \textbf{\bibinfo{volume}{74}}, \bibinfo{pages}{195304}
  (\bibinfo{year}{2006}).

\bibitem[{\citenamefont{Wu et~al.}(2010)\citenamefont{Wu, Yu, Li, Pey, Pan,
  Chai, Chiu, Lin, Xu, Wann et~al.}}]{wu:113510}
\bibinfo{author}{\bibfnamefont{L.}~\bibnamefont{Wu}},
  \bibinfo{author}{\bibfnamefont{H.~Y.} \bibnamefont{Yu}},
  \bibinfo{author}{\bibfnamefont{X.}~\bibnamefont{Li}},
  \bibinfo{author}{\bibfnamefont{K.~L.} \bibnamefont{Pey}},
  \bibinfo{author}{\bibfnamefont{J.~S.} \bibnamefont{Pan}},
  \bibinfo{author}{\bibfnamefont{J.~W.} \bibnamefont{Chai}},
  \bibinfo{author}{\bibfnamefont{Y.~S.} \bibnamefont{Chiu}},
  \bibinfo{author}{\bibfnamefont{C.~T.} \bibnamefont{Lin}},
  \bibinfo{author}{\bibfnamefont{J.~H.} \bibnamefont{Xu}},
  \bibinfo{author}{\bibfnamefont{H.~J.} \bibnamefont{Wann}},
  \bibnamefont{et~al.}, \bibinfo{journal}{Applied Physics Letters}
  \textbf{\bibinfo{volume}{96}}, \bibinfo{eid}{113510}
  (pages~\bibinfo{numpages}{3}) (\bibinfo{year}{2010}).

\bibitem[{\citenamefont{Charbonnier et~al.}(2010)\citenamefont{Charbonnier,
  Leroux, Cosnier, Besson, Martinez, Benedetto, Licitra, Rochat, Gaumer, Kaja
  et~al.}}]{Charbonnier_TED_2010}
\bibinfo{author}{\bibfnamefont{M.}~\bibnamefont{Charbonnier}},
  \bibinfo{author}{\bibfnamefont{C.}~\bibnamefont{Leroux}},
  \bibinfo{author}{\bibfnamefont{V.}~\bibnamefont{Cosnier}},
  \bibinfo{author}{\bibfnamefont{P.}~\bibnamefont{Besson}},
  \bibinfo{author}{\bibfnamefont{E.}~\bibnamefont{Martinez}},
  \bibinfo{author}{\bibfnamefont{N.}~\bibnamefont{Benedetto}},
  \bibinfo{author}{\bibfnamefont{C.}~\bibnamefont{Licitra}},
  \bibinfo{author}{\bibfnamefont{N.}~\bibnamefont{Rochat}},
  \bibinfo{author}{\bibfnamefont{C.}~\bibnamefont{Gaumer}},
  \bibinfo{author}{\bibfnamefont{K.}~\bibnamefont{Kaja}}, \bibnamefont{et~al.},
  \bibinfo{journal}{IEEE Transactions on Electron Devices}
  \textbf{\bibinfo{volume}{57}}, \bibinfo{pages}{1809 } (\bibinfo{year}{2010}).

\end{thebibliography}

\end{document}